\documentclass{amsart} 
\usepackage{amsmath}
  \usepackage{graphics} 
  \usepackage{epsfig} 
\usepackage{graphicx}  
\usepackage{epstopdf}
 \usepackage{paralist}

\theoremstyle{definition}

\newtheorem{remark}{Remark}

\newcommand{\eps}{\varepsilon}

\title[Critical transitions in a model of a genetic regulatory system]
      {Critical transitions in a model of a genetic regulatory system}

\author[Jesse Berwald and Marian Gidea]{}


 \email{jberwald@ima.umn.edu}
 \email{Marian.Gidea@yu.edu}

\thanks{The first author is supported by the MCRN.  The second author is supported by  NSF grant  DMS-1201357, by   the National Cancer Institute grant P20 CA157069, by The Fund for Math of the IAS, and  by the MCRN}

\begin{document}
\maketitle

\centerline{\scshape Jesse Berwald}
\medskip
{\footnotesize
 \centerline{Institute for Mathematics and Its Applications}
   \centerline{ Minneapolis,  Minneapolis, MN 55455, USA}
} 

\medskip

\centerline{\scshape Marian Gidea}
\medskip
{\footnotesize
 \centerline{Yeshiva University, Department of Mathematical Sciences, }
   \centerline{New York, NY 10016 }
   \centerline{and}
\centerline{School of Mathematics, Institute for Advanced Study}
\centerline{Princeton, NJ 08540, USA }
}

\bigskip

 \centerline{(Communicated by the associate editor name)}

\begin{abstract}
We consider a  model for  substrate-depletion oscillations in genetic systems, based on  a stochastic differential equation  with a slowly evolving external  signal. We show the existence of critical transitions in the system. We   apply two methods to numerically test   the synthetic time series generated by the system for early indicators of critical transitions: a detrended fluctuation analysis method, and  a novel method based on topological data analysis (persistence diagrams).
\end{abstract}


\section{Introduction}\label{section:introduction}
Gene expression is the process by which the genetic code is used to
synthesize functional gene products (proteins, functional RNAs). The
timing and the level of gene production is specified by a wide range
of mechanisms, termed gene regulation. It is believed that a
significant number of genes express cyclically, with about $10-15$\%
of genes directly regulated by the circadian molecular clock.  Such gene
expression oscillations allow for rapid adaptation to changes in
intracellular and environmental conditions.  In general, the phase and
 amplitude of gene expression depend on the function of the gene and 
internal and external stimuli.  It has been postulated that gene
expression oscillation is a basic property of all genes, not
necessarily connected with any specific gene function
\cite{Ptitsyn2007}.

Gene regulatory systems are intrinsically stochastic. Stochasticity
originates in the statistical uncertainty of the chemical reactions
between molecules, and is inversely proportional to the square root of
the number of molecules. Thus, lower numbers of interacting molecules
yield increasingly significant statistical fluctuations. Besides
intrinsic stochasticity, gene expression is also subjected to
extrinsic stochasticity, due to environmental effects.  In general,
stochastic fluctuations are seen as a source of robustness and stability,
but sometimes can adversely affect a cell function \cite{Leier2006}.

A motivation for the modeling and simulation of gene regulatory systems
comes from synthetic biology, an emerging field of research devoted to
the design and construction of biological systems, based on
engineering principles.  An interesting analogy in \cite{Tyson2003}
compares a molecular network to an electrical circuit, where, instead
of resistors, capacitors and transistors connected via circuits one
has genes, proteins, and metabolites connected via chemical reactions
and molecular pathways. 
Some milestone achievements in this direction have been reported,
e.g., in \cite{Bulter2004,Elowitz,Gardner}.
Relatedly, one would also like to achieve a clear understanding of the
functionality of the molecular circuits and networks through
measurements of various product outputs, in the same way one
understands an electrical circuit through measurements of current,
voltage, and resistance. In particular, one would like to discern
possible ways in which systems subjected to varying parameters and
noise may switch between different stable regimes, or more generally,
between potential attractors.

It is the concept of suddenly shifts amongst stable regimes which we
explore in this paper. We investigate the occurrence of these {\em
  critical transitions} in a model of a genetic regulatory system. By
a critical transition we mean a sudden change of a system from one
stable regime (fixed point, limit cycle) to an unstable regime,
possibly followed by some other stable regime.  We consider a simple
genetic circuit that exhibits an oscillatory regime, and we study the
behavior of the system under noise.  Explicitly, we consider a
two-gene model whose oscillations depend on several parameters. We
show that the system undergoes a critical transition under slow
parameter drift. We accomplish this by recording the time series
generated by this model and analyzing the critical transitions. We
utilize numerical methods to identify early warning signals indicative
of critical transitions in the synthetic data.

First, we apply a statistical method, based on detrended fluctuation
analysis, to analyze these time series data. The method is described
in detail in Subsection \ref{subsection:DFA}. The tests are performed
using a windowed analysis of the data and reveal that the
autocorrelation of the time series increases towards $1$, and the
variance of the time series distribution grows steadily prior to a
critical transition. These signs are consistent with early warning
indicators of critical transitions described by others; for instance,
see results in Scheffer, {\em et al}~\cite{Scheffer2009}.

Secondly, we apply a method from topological data analysis, based on
{\em persistence diagrams}, which we describe in more detail in
Subsection \ref{subsection:persistence}. Again, we consider windows of
data from the time series. In this case, we consider these windows as
strings of data points to which we associate filtrations of Rips
complexes and for which we generate associated persistence diagrams
to analyze the topology of the data at different resolutions. The
persistence diagrams 
reveal qualitative changes in the topology of the strings of data
points prior to the critical transitions: the distribution of data
points becomes more widespread and/or asymmetric. While the detrended
fluctuation analysis has been previously used for detection of
critical transitions, the application of persistence diagrams, a
method from topological data analysis, is novel.

We also make a comparison between the two methods. While the detrended
fluctuation analysis introduces artificial choices and possible bias
(see also \cite{Bryce2012}), the proposed topological method is
inherently robust. Note that both methods can be applied to detect
or predict critical transitions in experimental data as well. As such, the
results from model testing may serve as benchmarks for testing data measured from real world sources.


\section{Background}
In this section we briefly describe critical transitions in the context of fast-slow systems. We consider both deterministic and  stochastic systems. Then we present a simple genetic regulatory model of the substrate-depletion oscillator type.

\subsection{Critical transitions}\label{sub:critical}
A fast-slow system of ordinary differential equations is a system of the type
\begin{eqnarray}
       \label{eqn:f}   x' &=& f(x,y), \\
       \label{eqn:s}\eps y' &=&g(x,y),
\end{eqnarray}
where $(x,y)\in\mathbb{R}^m\times\mathbb{R}^n$, $f:\mathbb{R}^{m+n}\to\mathbb{R}^m$, $g:\mathbb{R}^{m+n}\to\mathbb{R}^n$ are $C^{r}$-functions with $r\geq 3$, $\eps>0$ is a small parameter, and
${}'=\frac{d}{dt}$. One can regard $y$ as a fast variable.
Rescaling the time   $t=\eps\tau$ yields
\begin{eqnarray}
       \label{eqn:ftau} \dot x &=&\eps f(x,y), \\
       \label{eqn:stau} \dot y &=& g(x,y),
\end{eqnarray}
where $\,\dot{}= \frac{d}{d\tau}$. The singular limit of \eqref{eqn:f},\eqref{eqn:s} when $\eps\to 0$ gives the slow subsystem, and the singular limit of \eqref{eqn:ftau},\eqref{eqn:stau}  gives the fast subsystem. The critical set is defined  as
\[C_0=\{(x,y)\,:\,g(x,y)=0\}\] and consists of equilibrium points for the fast subsystem. If the Jacobian
$\frac{\partial g}{\partial y}$  is nonsingular on $C_0$, then $C_0$ is an $m$-dimensional manifold,   and is the graph of a smooth
function $y=h_0(x)$.  The slow subsystem is determined by $x'=f(x,h_0(x))$ and restricts to $C_0$. In the case that all eigenvalues of $\frac{\partial g}{\partial y}$ at a point have non-zero real parts, then the point is normally hyperbolic. The set of normally hyperbolic points forms a normally hyperbolic invariant manifold (NHIM). In particular, it has stable and unstable manifolds. The NHIM can have attractive components, where all the eigenvalues of  $\frac{\partial g}{\partial y}$  have negative real part, and repelling components, where at least one eigenvalue has positive real part.
An attractive component has an $n$-dimensional stable manifold, while a repelling component has a non-trivial unstable manifold.

Fenichel's Theorem \cite{Fenichel79} implies that, for all sufficiently small $\eps$, every compact submanifold (with boundary) $S_0$ of $C_0$ can be continued to a NHIM $S_\eps$ (not uniquely defined) for the flow of \eqref{eqn:f}, \eqref{eqn:s},  which is the graph of a smooth
function $y=h_\eps(x)$.
The stable and unstable manifolds of $S_0$ continue to stable and unstable manifolds of $S_\eps$. The flow on $S_\eps$ converges to the slow flow as $\eps\to 0$. Such a manifold $S_\eps$ is referred to as a slow manifold.

We define a critical transition for this type of system following \cite{Kuehn2011}. We assume that the critical set $C_0$ can be decomposed as  $C_0=S^a_0\cup S^r_0\cup S^b_0$, where $S^a_0$ is an attractive NHIM, $S^r_0$  a repelling NHIM, and $S^b_0$ is the part of $C_0$ that is not normally hyperbolic (corresponding to bifurcation points). By definition, a point $p_0=(x_0,y_0)$  on $C_0$ that is not normally hyperbolic is a critical transition if there exists a concatenation of trajectories $\gamma_0$, $\gamma_1$, where $\gamma_0:[t_{-1},t_0]\to \mathbb{R}^{m+n}$,  $\gamma_1:[t_0,t_1]\to \mathbb{R}^{m+n}$ satisfy the following properties:
\begin{itemize}
  \item[(1)] $\gamma_0(t_{-1},t_0)$ is a trajectory of the slow subsystem, oriented from $\gamma_0(t_{-1})$ to $\gamma_0(t_0)$, contained in the attracting NHIM $S^a_0$;
  \item[(2)] $\gamma_0(t_0)=\gamma_1(t_0)=p_0\in S^b_0$ is a point that is not normally hyperbolic;
  \item[(3)] $\gamma_1(t_0,t_1)$ is a trajectory of the fast subsystem, oriented from $\gamma_1(t_0)$ to $\gamma_1(t_1)$.
\end{itemize}

When we consider the dynamics of the system for $\eps>0$ sufficiently
small, a trajectory starting near $\gamma_0$ follows closely the slow
dynamics around $\gamma_0$ for some time, after which it transitions
to follow the fast dynamics near $\gamma_1$ for a period of time. In
\cite{Kuehn2011,Kuehn2012} several types of bifurcations are examined
to determine whether or not they exhibit the characteristics of
critical transitions, under some suitable conditions on the
smoothness, compactness, and non-degeneracy on the system. We
summarize the findings below:
\begin{itemize}
\item[(a)] for $m=n=1$, saddle-node (fold) bifurcation points determine critical transitions;
\item[(b)] for  $m=n=1$, subcritical pitchfork bifurcation points determine critical transitions;
\item[(c)] for  $m=n=1$, transcritical bifurcation points determine critical transitions;
\item[(d)] for  $m=2$ and $n=1$,   subcritical non-degenerate Hopf bifurcations determine critical transitions.
\end{itemize}

\subsection{Incorporating noise}\label{sec:noise}

It is often important for the understanding of a physical system to
incorporate stochastic effects. We consider the Langevin form of
\eqref{eqn:f}, \eqref{eqn:s}
\begin{eqnarray}
       \label{eqn:fst}   dx &=& f(x,y)+\sigma_1dW_1, \\
       \label{eqn:sst}  dy&=&\frac{1}{\eps}g(x,y)+\frac{1}{\sqrt{\eps}}\sigma_2dW_2,
\end{eqnarray}
where $\sigma_1$, $\sigma_2$ represent noise levels (depending on
$\eps$), and $W_1,W_2$ are one-dimensional Wiener processes (Brownian
motions).  Assuming $\sigma_1,\sigma_2$ are sufficiently small, the
sample paths of the system \eqref{eqn:fst}, \eqref{eqn:sst} stay near
$S^a_0$ with high probability, up to a neighborhood of the critical
transition, after which they exit the
neighborhood. In~\cite{Kuehn2011,Kuehn2012}, it is argued that the following
behaviors are typical of a system prior to a critical
transition:
\begin{itemize}
\item[(i)] The system recovery  from small perturbations is `critically' slows down;
\item[(ii)]  The variance in the time series  increases steadily;
\item[(iii)]   The autocorrelation of the time series increases towards $1$;
\item[(iv)]   The distribution of the time series becomes more asymmetric.
\end{itemize}

We note that (iv) from above depends on whether or
not the underlying bifurcation has symmetry. For example, a
saddle-node bifurcation as in (a) from above will typically yield
asymmetric fluctuations, while a pitchfork bifurcation as in (b) from
above will typically yield symmetric fluctuations. A complementary
characteristic to (iv) is that the distribution of the time series
loses its normality, for example it changes from uni-modal to
multi-modal.

Such system response characteristics can be monitored
numerically and serve as `early warnings' of critical transitions
in real-world systems. Examples include Earth's climate, ecological
systems, global finance, asthma attacks or epileptic seizures; see,
e.g., \cite{Ditlevsen2010,Scheffer2009,Scheffer2012,Thompson2010}, and
the references therein.

\subsection{A simple genetic circuit}
We briefly describe a model of simple genetic circuit which generates
oscillations of varying amplitude.  The model consists of two genes,
one producing the protein $R(t)$, and the other producing the protein
$X(t)$.  The protein $X(t)$ is a substrate for the activator protein
$R(t)$ that is produced in an autocatalytic process. As $R(t)$
accumulates, the production of $R(t)$ accelerates until there is an
explosive conversion of the whole of $X(t)$ into $R(t)$. This rapid
change corresponds to a critical transition in the underlying system.
With the substrate $X(t)$ depleted, the autocatalytic reaction
terminates, and the activator $R(t)$ degrades in time.  This allows
the level of $X(t)$ to grow again, leading to another cycle of
explosive growth in $R(t)$.  This process is know as a
substrate-depletion oscillator.

An example of this mechanism is the oscillation of the
M-phase-promoting factor (MPF) activator in the frog egg, where the
substrate is the phosphorylated form of the B-cyclin-dependent kinase
(note that the true mechanism involves several other proteins and
reactions)~\cite{Novak1993}. A mathematical model for the
substrate-depletion oscillator is given by the following system:
\begin{eqnarray}
       \label{eqn:X}  X'(t) &=& k_1S -[k_0'+k_0E_P(R(t))]X(t), \\
       \label{eqn:R} R'(t) &=&[k_0'+k_0E_P(R(t))]X(t)-k_2R(t).
\end{eqnarray}
Here $E_P(R(t))$ represents the level of the phosphorylated version of
the protein $R(t)$ -- involved with $R(t)$ in a mutual activation
process -- given by $E_P(R)=G(k_3R,k_4,J,K)$, where $G$, the Goldbeter-Koshland function, is defined by
\[G(u,v, J,K)=
\frac{2uK}{v-u+vJ+uK+\sqrt{(v-u+vJ+uK)^2-4(v-u)uK}}.\] 
The Goldbeter-Koshland function represents the equilibrium
concentration of the phosphorylated form of a protein, for a
phosphorylation-dephosphorylation reaction governed by Michelis-Menten
kinetics. The Goldbeter-Koshland function is responsible for creating
a switch-like signal-response in the evolution of the protein $R(t)$.
The quantities $k_0,k'_0,k_1,k_2,k_3,k_4,J,K,S$ are parameters. The
parameter $S$ is the strength of a signal, representing the rate of
synthesis of the substrate $X$, which we regard as an external input
to the system.

This system presents both positive and negative feedback. The positive
feedback loop creates a bistable system and the negative-feedback loop
drives the system back and forth between two stable steady states.  In
what follows, we will modify the simple genetic circuit
in~\eqref{eqn:X} and~\eqref{eqn:R} by considering the external input
$S$ as a slowly varying parameter, in addition to including a
stochastic term. We will study critical transitions in the resulting
system.

\section{Model}\label{section:model} We consider a new model of a genetic regulatory network  with a slowly dependent signal, given by
\eqref{eqn:X}, \eqref{eqn:R}, with $S$ being now a slowly evolving
parameter, i.e.
\begin{eqnarray}
       \label{eqn:XS0}  X'(t) &=& k_1S(t) -[k_0'+k_0E_P(R(t))]X(t), \\
       \label{eqn:RS0} R'(t) &=&[k_0'+k_0E_P(R(t))]X(t)-k_2R(t),\\
       \label{eqn:S0} S'(t)&=&\eps,
\end{eqnarray}
where $\eps>0$ is small.
We will fix the parameter as in \cite{Tyson2003},
$k_0=0.04,k'_0=0.01, k_1=k_2=k_3=1,k_4=0.3,J=K=0.05$.

The fast subsystem is obtained by letting $\eps\to 0$ yielding
\begin{eqnarray}
       \label{eqn:Xf}  X'(t) &=& k_1S(t) -[k_0'+k_0E_P(R(t))]X(t), \\
       \label{eqn:Rf} R'(t) &=&[k_0'+k_0E_P(R(t))]X(t)-k_2R(t),\\
       \label{eqn:Sf} S'(t)&=&0.
\end{eqnarray}

\begin{figure}
\includegraphics[width=1\textwidth, clip, keepaspectratio]
{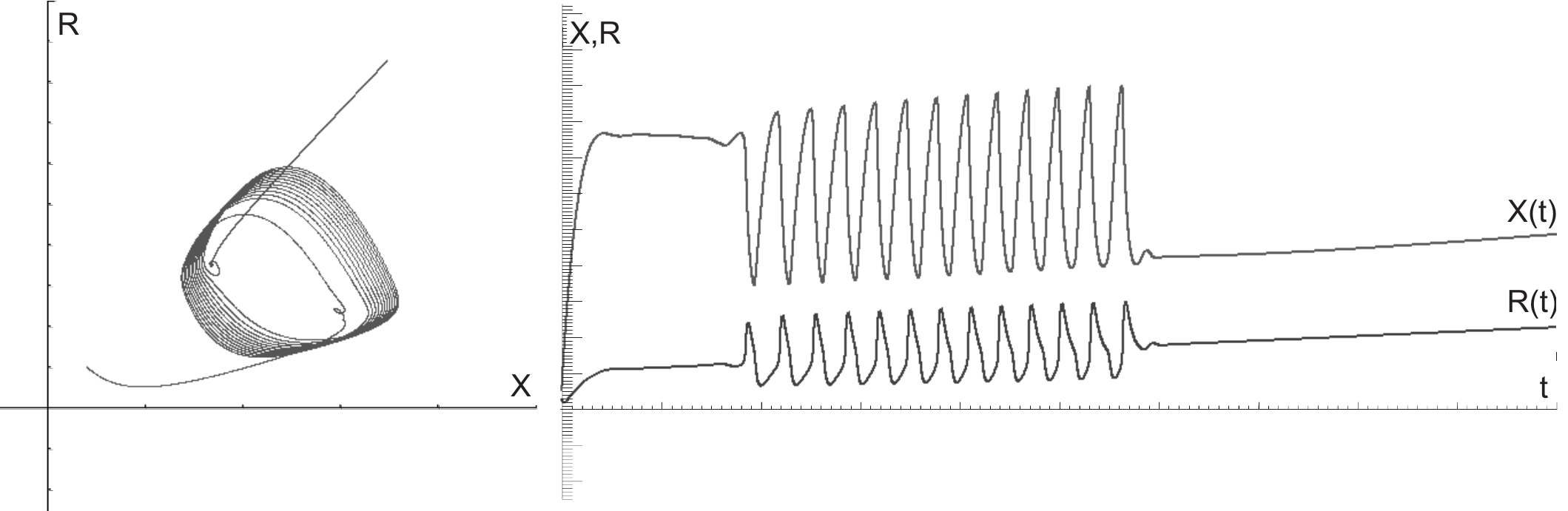}
 \caption[]{A sample trajectory, and the $X$-time series (higher value) and $R$-time series (lower values).}
    \label{sdevol}
\end{figure}

We rescale time $\tau=\eps t$, and we rewrite the corresponding system relative to the rescaled time
\begin{eqnarray}
     \label{eqn:Xr}  \eps\dot X &=& k_1S -[k_0'+k_0E_P(R)]X, \\
       \label{eqn:Rr} \eps\dot R  &=&[k_0'+k_0E_P(R )]X -k_2R ,\\
       \label{eqn:Sr} \dot S &=&1.
\end{eqnarray}
Sample trajectories for this system are plotted in Fig.~\ref{sdevol}. From the above equations the slow subsystem is obtained by letting $\eps\to 0$ yielding
\begin{eqnarray}
     \label{eqn:Xs}  0 &=& k_1S -[k_0'+k_0E_P(R)]X, \\
       \label{eqn:Rs} 0  &=&[k_0'+k_0E_P(R )]X -k_2R ,\\
       \label{eqn:Ss} \dot S &=&1.
\end{eqnarray}
From this we see that the critical submanifold is given by
\[C_0=\left\{(X,R,S)\,,\, X=\frac{k_1S}{k_0'+k_0E_P({\frac{k_1}{k_2}S)}}, R=\frac{k_1}{k_2}S\right\}.\]

The critical submanifold consists of equilibrium points of the fast subsystem. The stability of the $C_0$
is determined by the eigenvalues of the Jacobi matrix evaluated at the equilibrium points
\[J=\left(
      \begin{array}{cc}
        -[k'_0+k_0E_P(R)] & -[k_0\partial_RE_P(R)X] \\
        k'_0+k_0E_P(R) & k_0\partial_RE_P(R)X-k_2\\
      \end{array}
    \right).\]

For the values of the parameters chosen above, we find that the
stability at an equilibrium point changes at $S_{crit1}=0.13326703$
and $S_{crit2}=0.34680193$, respectively. The corresponding equilibria are
$X_{crit1}=5.41285587$, $R_{crit1}=0.13326703$, and
$X_{crit2}=1.07370450$, $R_{crit2}=0.34680193$.

\begin{figure}
$$\begin{array}{cc}
\includegraphics[width=0.5\textwidth, clip, keepaspectratio]
{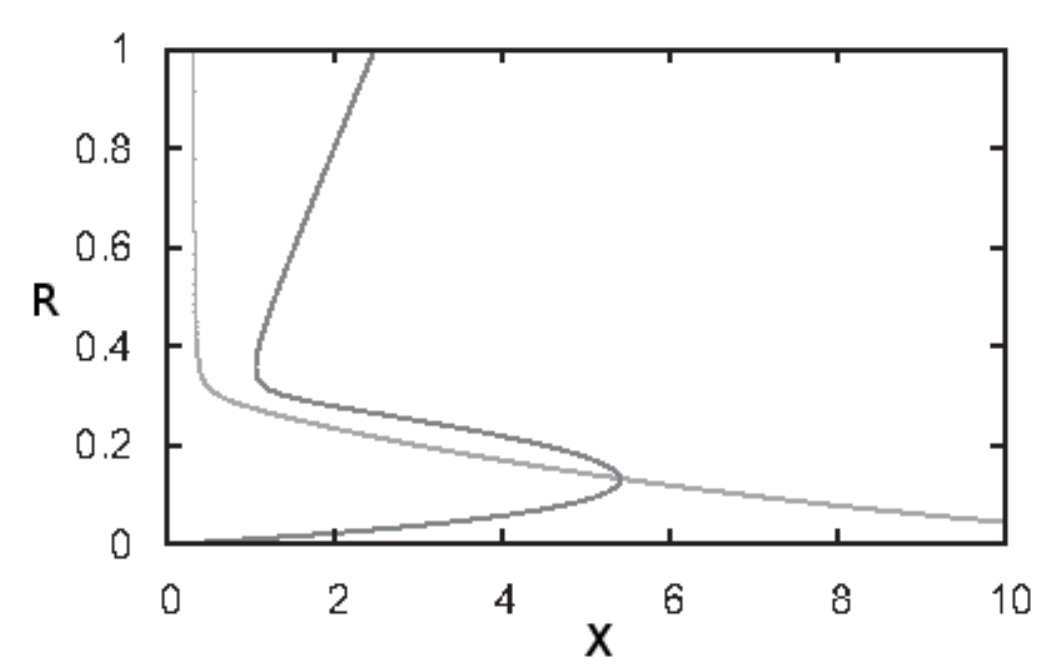} &
\includegraphics[width=0.5\textwidth, clip, keepaspectratio]
{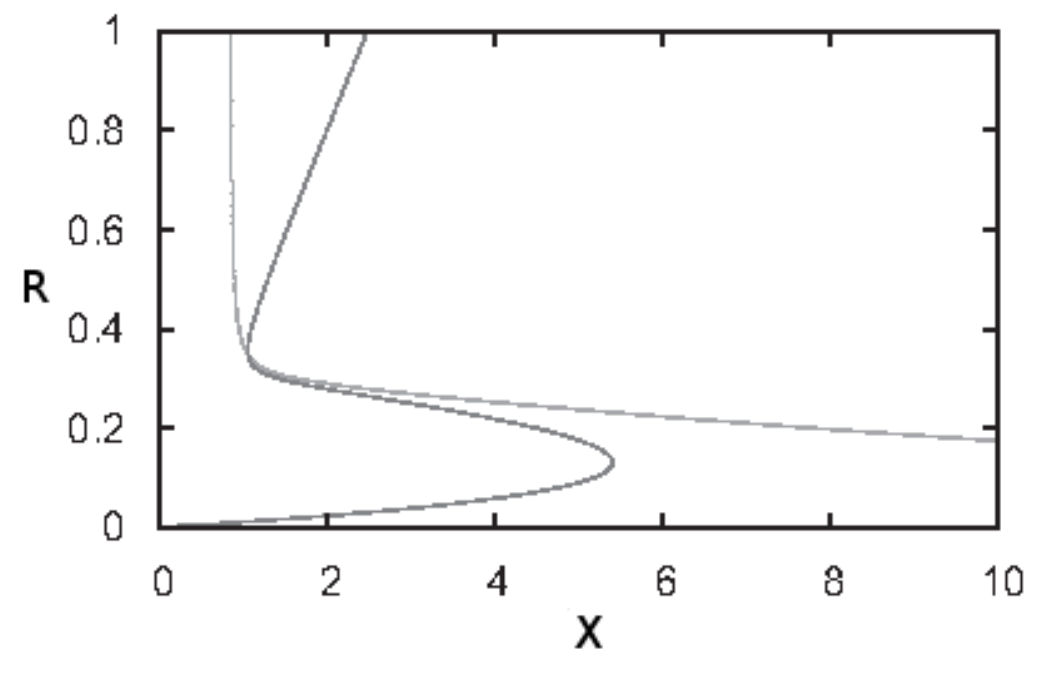}
\end{array}
$$    \caption[]{The nullclines of the fast system for $S=S_{crit1}$ and $S=S_{crit2}$.}
    \label{nullclines}
\end{figure}

Precisely, for $S<S_{crit1}$ and $S>S_{crit2}$ the equilibrium point
is stable. For $S\in(S_{crit1}, S_{crit2})$ the equilibrium point is
unstable, and there exists a periodic orbit that is asymptotically
stable, whose existence can be established numerically, as observed in Fig.~\ref{sdevol}.  The values
$S=S_{crit1}$, $S=S_{crit2}$ yield subcritical Hopf bifurcations,
where an unstable equilibrium point is turned into a stable one and a
small unstable periodic orbit is born (or vice versa).  In addition,
one expects canard-type solutions in some exponentially small
neighborhoods of $S_{crit1}$, $S_{crit2}$, relative to $\eps$ (see,
e.g., \cite{KrupaS2001}). In Fig.~\ref{nullclines} we plot the
nullclines of the system for the critical points $S_{crit1}$ and
$S_{crit2}$.


The specific stochastic differential equation (SDE) system associated to \eqref{eqn:Xr}, \eqref{eqn:Rr}, \eqref{eqn:Sr} can be written
\begin{eqnarray}
     \label{eqn:Xr_stoch}  \dot X &=&\frac{1}{\eps}(k_1S -[k_0'+k_0E_P(R)]X)+\frac{\sigma_1}{\sqrt{\eps}}dW_1, \\
       \label{eqn:Rr_stoch} \dot R  &=&\frac{1}{\eps}([k_0'+k_0E_P(R )]X -k_2R)+\frac{\sigma_2}{\sqrt{\eps}}dW_2,\\
       \label{eqn:Sr_stoch} \dot S &=&1,
\end{eqnarray}
where $W_1,W_2$ represent Brownian motions, and $\sigma_1,\sigma_2$
are noise levels. Since the parameter values $S_{crit1}$ and
$S_{crit2}$ yield subcritical Hopf bifurcations, the theory from
Subsection \ref{sub:critical} allows us to conclude that the
corresponding points $(X_{crit1}, R_{crit1})$, $(X_{crit2},
R_{crit2})$ determine critical transitions. A typical trajectory of
the stochastic system in the phase space and its corresponding $R$-time series
are shown in Fig.~\ref{substratedepletion}. Examining the $R$-time
series, one can see that a critical transition occurs near time
$t \approx 1000$. We note that a similar analysis for activator-inhibitor
oscillations has been performed in \cite{Kuehn2012}.

\begin{remark} As mentioned in Section \ref{section:introduction}, 
noise in the form of random fluctuations arises naturally in gene
regulatory networks.  One typically distinguishes between intrinsic
noise, inherent in the biochemical reactions, and extrinsic noise,
originating in the random variation of the externally set control
parameters.  Both types of noise can be model by augmenting the
governing rate equations with additive or multiplicative stochastic
terms. We refer the interested reader to~\cite{Hasty2000}.

\begin{figure}\centering
\includegraphics[width=1.0\textwidth, clip, keepaspectratio]{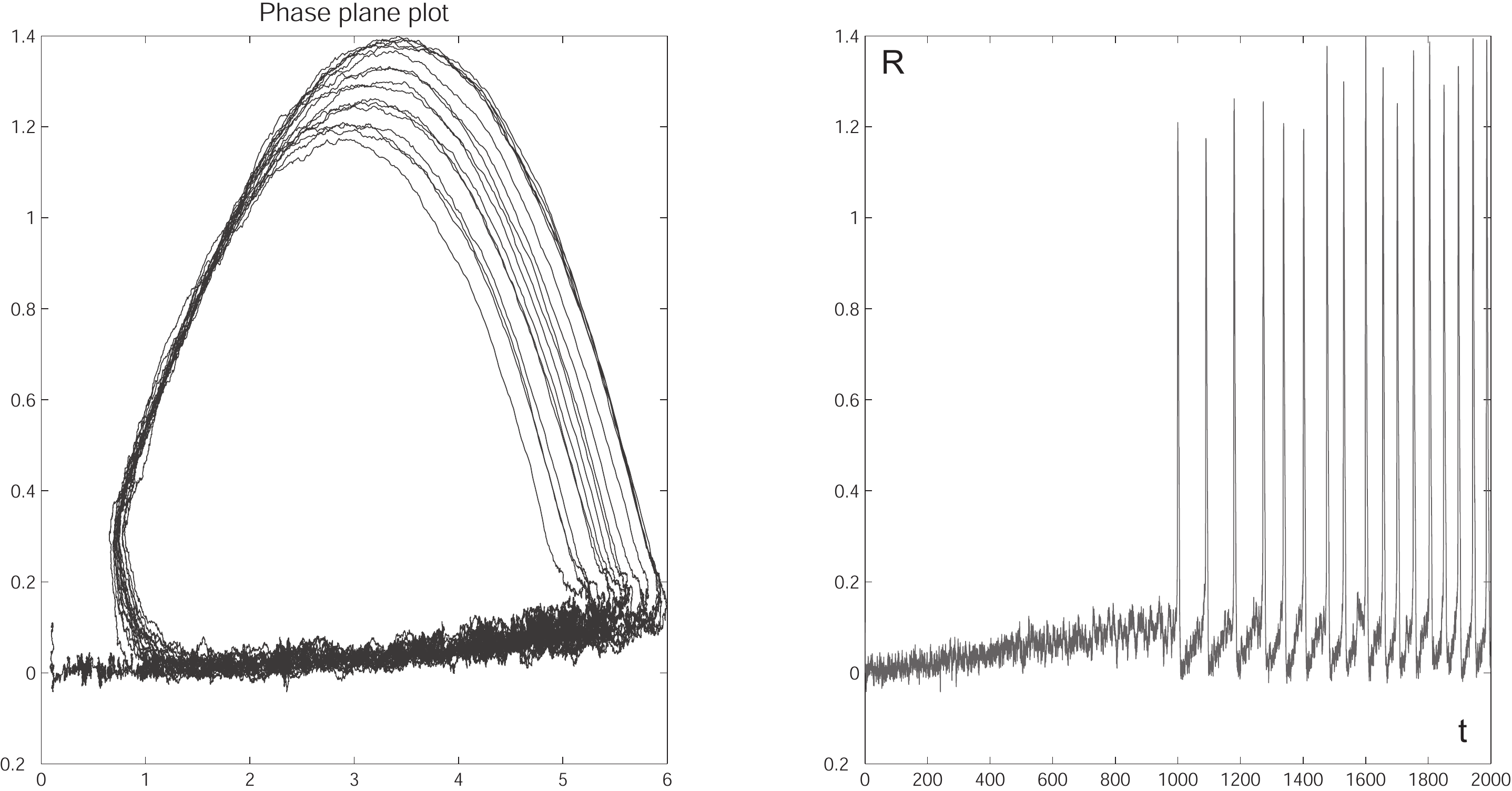}
\caption[]{Phase space of the model described by \eqref{eqn:Xr_stoch}, \eqref{eqn:Rr_stoch}, \eqref{eqn:Sr_stoch},    and a corresponding $R$-time series.}
    \label{substratedepletion}
\end{figure}

In our model we only consider the effects of additive noise, which can
be thought of as a randomly varying external field acting on the
biochemical reactions. The field enters into the governing rate
equations as an additive stochastic term in the Langevin equation.  We
choose to focus on additive extrinsic noise as this could be used as a
switch and/or amplifier for gene expression, which has potential
applications to gene therapy~\cite{Hasty2000}.  Switching mechanisms
are exactly the type of phenomena that we would like to capture via
the critical transitions approach.
\end{remark}

\section{Methods}\label{sec:methods}

We use the model proposed in Section~\ref{section:model}  to generate a
synthetic time series given by successive reading of one of the
variables. 
We investigate the synthetic time series for early warning signs of critical
transitions. Below we describe two such detection methods: a well-known
detrended fluctuation analysis method, and a novel method inspired by
topological data analysis.

\subsection{Detrended fluctuation analysis}\label{subsection:DFA}

Detrended fluctuation analysis (DFA) is a technique introduced by
Hurst half a century ago to analyze fluctuations in time series.  The
DFA procedure has been widely used for early detection of critical
transitions \cite{Livnia2007,Scheffer2009,Thompson2010}. We outline
the algorithm below.

\subsubsection{Algorithmic description of DFA} The DFA procedure takes as input a time series $(s_k,z_k),\, k=1,\ldots, N$, where $s_k$ is the instant of time of the $k$-th measurement (not necessarily equally spaced), and $z_k$ is the $k$-th measurement of some observable. To detect whether the system undergoes a critical transition, the DFA algorithms proceeds as follows:
\begin{description}
	 \item[Interpolation] Choose an optimal step size $\Delta t$,
           and interpolate the given time series such that it is
           evenly spaced in time. Denote the new series $(t_k, x_k)$,
           with $t_{k}=k\Delta t$.

         \item[Detrending] One way to remove a general trend from statistical
           data is by subtracting a moving average.  For
           example,  using a Gaussian kernel
           \[G_k(t)=\frac{1}{\sqrt{2\pi}d}\exp\left(-\frac{(t-k\Delta t)^2}{2d^2}\right)\]
           of bandwidth $d$, one may compute the weighted average of $x_k$,
\[X(k\Delta t)=\frac{\sum_{i=1}^{N}G_k(i\Delta t)x_i }{\sum_{i=1}^{N}G_k(i\Delta t)}.\]
Subtracting the weighted average from the time series
yields the detrended series,
\[ y_k=x_k-X(k\Delta t)\].

\begin{remark}
  Instead of Gaussian kernel detrending, alternative detrending
  methods can be used, such as linear, cubic spline, or Fourier
  interpolation, depending on the nature of the data. We also explore
  these additional methods in Section~\ref{sec:results}
\end{remark}

  \item[Lag-1 autocorrelation] 
    The final step involves fitting a first-order autoregressive (AR(1)) process
      \[y_{k+1}=c_ky_k+\sigma \xi_k,\]
 to the detrended time series $y_k$, where $(\sigma
 \xi_k)_{k\in\mathbb{N}}$ is white noise of intensity $\sigma$. In
 order to compute $c_k$, choose a sliding window of size $w$ and determine the
 least-squares fit
\[y_{j+1} \approx c_ky_j,\quad \textrm{ for } j=k,\ldots, k+w-1.\]
Hence, for each window we extract the value of $AR(1)$ $c_k$. Recall that the lag-1
autocorrelation $AR(1)$ is $0$ for white noise and close to $1$ for
red (autocorrelated) noise.
\end{description}

\subsubsection{Detection of critical transitions via DFA}\label{sec:dfa-detection}

The following criterion has been proposed for the detection of critical transition \cite{Scheffer2009}:
\begin{itemize}
  \renewcommand{\labelitemi}{$\vcenter{\hbox{\tiny$\bullet$}}$}
\item Given a time series measured from a system approaching a
  critical transition, the DFA outputs a time series $(t_k,y_k)$ for
  which
\begin{enumerate}
  \item the autocorrelation has a general trend which increases towards $1$;
  \item the variance has a positive trend.
\end{enumerate}
\end{itemize}
In certain special cases, this criterion has a  rigorous justification~\cite{Kuehn2011,Kuehn2012}.

The DFA method is an effective tool in detecting early signs of
critical transitions in noisy data. However, the method comes with
several significant drawbacks, such as its sensitivity to the
procedures and parameters used in processing the data. For instance,
the sample frequency, detrending method (e.g., the bandwidth of the
Gaussian detrending), or the size of the sliding window all have a
strong effect on the conclusions drawn from the algorithm in
subsection~\ref{subsection:DFA} and hence on the power of the DFA
method to serve as a prediction tool. One concern is that the
measurement of the $AR(1)$ values as well as the variance are strongly
influenced by the fit of the detrending method, with a poor fit being
likely to signal `false positives' for critical transitions (see
\cite{Bryce2012}).

\subsection{Persistence diagrams}\label{subsection:persistence}

As mentioned in Section~\ref{section:introduction}, we propose to use
tools from the field of topological data analysis as a new method to
detect critical transitions in dynamical systems.  In particular, we
leverage the stability properties of persistence diagrams to detect
critical transitions.  Topological persistence is a relatively recent
development that forms the core of topological data analysis and has
been widely used to extract relevant information from noisy data (see
\cite{Edelsbrunner2008} for background in persistence topology
in general).  There are numerous applications, including computer vision,
cluster analysis, biological networks, cancer survival analysis, and
granular material (see
\cite{Chazal2011,Edelsbrunner2008,Gameiro2012,Kondic2012,Nicolau2011}
and the references listed there).

In this section we describe the way in which we adapt this method to
observe changes in time series from systems approaching or undergoing
critical transitions. The key idea is to extract from the time series
consecutive strings of data points of a fixed length, which we regard
as individual point cloud data sets.  To each such point cloud we
assign a topological invariant, namely its persistence
diagram. Roughly speaking, the persistence diagram is a representation
of the data set in an abstract metric space which encodes information
about topological features of the data.

The highlight of this method is that when the system undergoes a
critical transition, the topological features associated to the
point cloud data sets also change significantly. The fact that the
corresponding persistence diagrams and distances between them can be
computed algorithmically enables us to describe these changes quantitatively.

\subsubsection{Description of persistence diagrams}
We describe the concept of a persistence diagram associated with point
cloud data starting with an informal description. From a high-level
perspective, the data analysis pipeline works as follows:
\[ \text{Data } 
\implies \text{ Filtration } 
\implies \text{ Persistence Module } 
\implies \text{ Persistence Diagram}
\]
We focus on the first two and the fourth parts of this pipeline, and only 
briefly detail the algebraic aspects of the third component below.
Suppose that one is provided with a point cloud data set, $X_0$, that
is an approximation of some geometric shape. One would like to infer
from the data the topological information on that shape. However, a
finite collection of points has only trivial topology. One way to
convert the collection of points into a non-trivial topological space
is to replace the points of the set by balls of a certain radius
$\epsilon$. One then computes the topological features of the
resulting set, $X_{\epsilon}$. Typical invariants resulting from this
computation, which serve to classify the set, include the number
connected components along with the number of `tunnels' and number of
`cavities' (known as Betti numbers).

\begin{figure}\centering
\includegraphics[width=0.7\textwidth, clip, keepaspectratio]{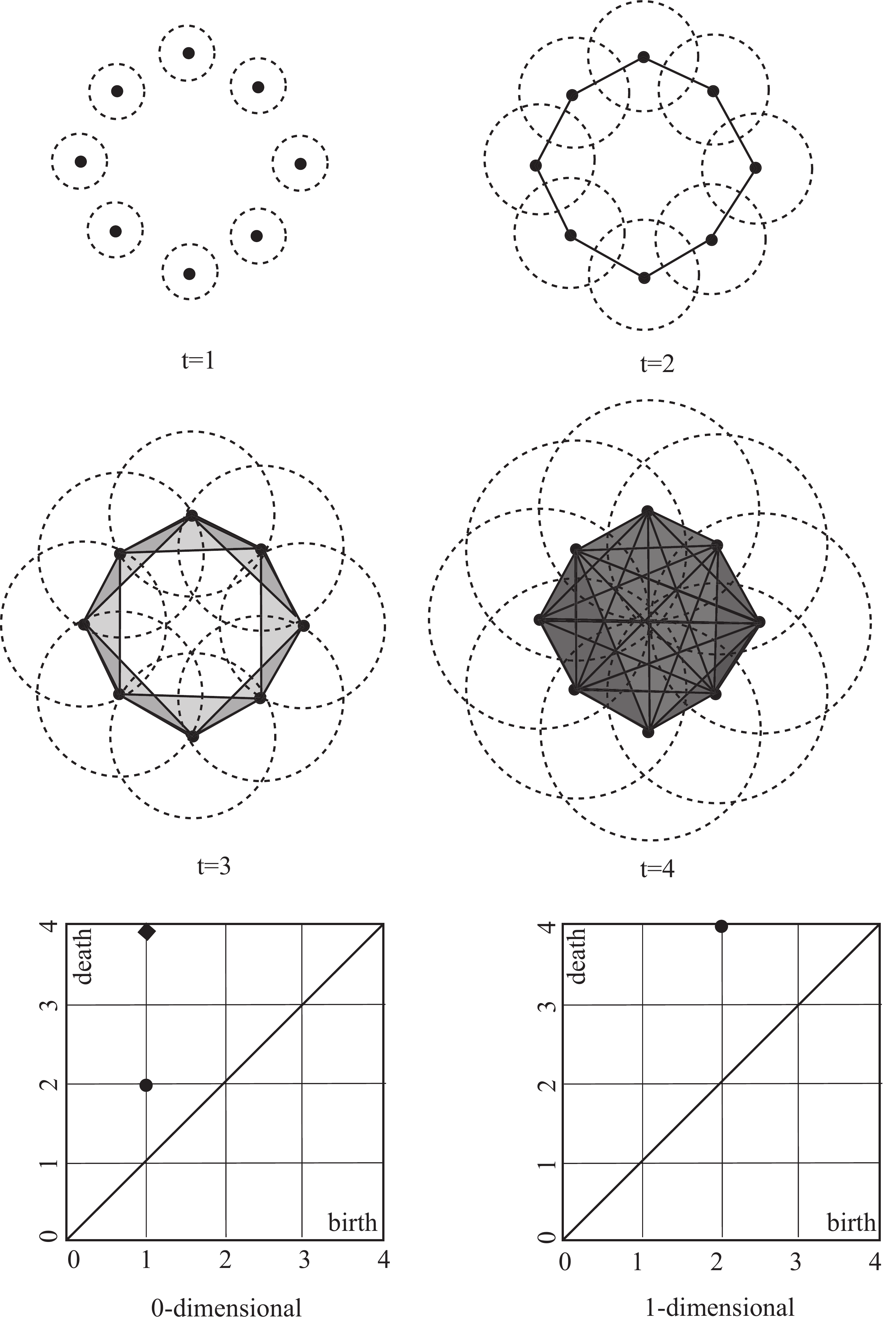}
\caption[]{A collection of points in the plane, resembling a `noisy'
  circle, is given.  At each instant of time $t=1,\ldots,4$, around
  each point we constructs disks of radii $\delta_1/2,\ldots,
  \delta_4/2$, respectively, with
  $\delta_1<\delta_2<\delta_3<\delta_4$.  The corresponding Rips
  complexes are also constructed at each instant.  The topological
  features of the Rips complexes change as time increases.  At time
  $t=1$ there are 8 connected components and no 1-dimensional hole. At
  time $t=2$ the connected components coalesce into a single component
  (thus in the 0-dimensional diagram there are actually 7 deaths
  represented at $(1,2)$), and a 1-dimensional hole is born.  Both the
  single connected component and the 1-dimensional hole survive to
  $t=3$. At time $t=4$ the 1-dimensional hole dies as it fills in (the
  death is represented at $(2,4)$), while the single connected
  component continues living (in fact, it has infinite lifespan). The
  $\Diamond$ at $(1,4)$ represents this
  fact. 
}
    \label{persistence}
\end{figure}

Of course, the topology of $X_{\epsilon}$ depends on the choice of the
radius of the balls in this construction. Instead of fixing a certain
radius, topological persistence considers all possible radii, from
some sufficiently small value, up to a sufficiently large radius. This
growth of the radius yields the filtration step above. As the radius
is gradually increased, new topological features will be `born' and
certain existing ones will `die'. A schematic representation of this
process is depicted in Fig.~\ref{persistence}.  The birth and death of
each topological feature at a given dimension is recorded by a
persistence diagram. This is a collection (multiset actually) of
(birth,death) times in $\mathbb{R}^2$. The 0- and 1-dimensional
diagrams for the associated filtration are shown in the bottom row of
Fig.~\ref{persistence}. 

The lifespan of a feature is easily computed
by calculating (death time) - (birth time). A topological feature with
a long lifespan, measured by the range of radii over which it
`persists',
is likely to capture an essential topological feature of the
underlying space from which the data was sampled. On the other hand,
short lived features are likely to result from `noise' in the
data. However, rather than discriminating between what is an essential
feature of the topology and what is not, the persistence diagram
method provides a summary of topological features that appear and
disappear throughout the variation of the radii of the balls, as well
as a ranking of the significance of these features, expressed in terms
of the lifespans.

We now continue with a concise, formal description of persistence
diagrams. For an introduction to algebraic topology and homology,
see~\cite{Hatcher2002}; surveys of persistent topology can be found
in~\cite{Edelsbrunner2008,Zomorodian2005}
 Given point cloud data $X\subset \mathbb{R}^d$, i.e., a collection of points in $\mathbb{R}^d$, and $\delta>0$, we associate to them the Rips complex $\mathcal{R}_\delta$. This is, by definition,   the abstract simplicial complex whose $0$-simplices are points $x_\alpha$, and whose   $k$-simplices are given by unordered $(k+1)$-tuples of points $\{x_{\alpha_j}\}_{j=0,\ldots,k}$ which are pairwise within a distance $\delta$. Figure \ref{persistence} provides an example for $X \subset \mathbb{R}^2$.
For all $0<a<b$ we  have $\mathcal{R}_a\subset\mathcal{R}_b$.  That is, the family $\{\mathcal{R}_\delta\}_{\delta>0}$ forms a filtration.

Denote by $H_p(\mathcal{R}_a)$ the $p$-homology of $\mathcal{R}_a$
with $\mathbb{Z}_2$ coefficients.  Heuristically, the homology of a
simplicial complex provides information about the topological features
of the complex, e.g., the number of connected components, tunnels and
cavities in that complex. The inclusion $\mathcal{R}_a\hookrightarrow
\mathcal{R}_b$ induces the homomorphisms $f^{a,b}_*:H_*(\mathcal{R}_a) \to H_*(\mathcal{R}_b)$ in all dimensions. Note that the image $F^{a-\rho,b}$ of $f^{a-\delta,a}_*$ in $H_*(\mathcal{R}_b)$ is independent of $\rho$ for all $\rho>0$ sufficiently small. We denote this image by $F^{a-,b}_*$.

A real value $c>0$ is called a homological critical value if there exists $q$ such that the homomorphism $f^{c-\rho,c}_q:H_q(\mathcal{R}_{c-\rho}) \to H_q(\mathcal{R}_{c})$ is not an isomorphism for all sufficiently small $\rho>0$. The image $F_q^{c-,c}$ of $f^{c-\rho,c}_q$ in $H_q(\mathcal{R}_{c})$ is independent of $\rho$, if this is small enough.
The quotient group $B_q^c=H_q(\mathcal{R}_c)/F_q^{c-,c}$ is called the $q$-th birth group at $\mathcal{R}_c$, and it captures the homology classes that did not exist in $\mathcal{R}_{c-\rho}$. A homology class $\alpha \in H_q(\mathcal{R}_c)$ is born in $\mathcal{R}_c$ if it represents a non-trivial element in $B_q^c$, that is, the canonical projection of $\alpha$ is non-zero.

Now consider the homomorphism $g^{a,b}_q:B_q^a\to H_q(\mathcal{R}_b)/F^{a-,b}$, where $g^{a,b}_q([\alpha])=[f^{a,b}_q(\alpha)]$, for $\alpha \in H_q(\mathcal{R}_b)$, where the notation $[\cdot]$ stands for equivalence class. We set $g_q^{a,b}=0$ for all $b>0$ sufficiently large. The kernel $D_q^{a,b}$ of the map $g^{a,b}_q$  is called the death subgroup of $B_q^a$ at $\mathcal{R}_q^b$. A homology class   $\alpha\in H_q(\mathcal{R}_a)$ dies entering $\mathcal{R}_b$ if $[\alpha]\in D_q^{a,b}$ but $[\alpha]\not\in D_q^{a,b-\rho}$, for $\rho>0$ sufficiently small. The degree $r$ of the death value $b$ of $B^a_q$ is  defined by $r=\textrm{rank}D_q^{a,b}-\textrm{rank}D_q^{a,b-}$. The sum of the degrees of all death value of the birth group $B_q^a$ is clearly equal to $\textrm{rank}(B^a_q)$.
The birth time of a homology class $\alpha$ is the value $a>0$ where the $\alpha$ is born in $\mathcal{R}_a$, and
the death time is the value $b>0$ where   $\alpha$ dies in $\mathcal{R}_b$.

The $q$-persistence diagram of the filtration $(\mathcal{R}_\delta)_{\delta>0}$ is defined as a multiset  $\mathcal{P}_q$ in $\mathbb{R}^2$ consisting of points of the type $z_i=(a,b_i)$, where $a$ is a birth value corresponding to a non-trivial group $B_p^a$, and $b_i$ is a death value of  $B_p^a$; the point $z_i$ appears in the diagram with multiplicity equal to the degree $r_i$ of the death value $b_i$. Since deaths occur after births, all points $(a,b_i)$ lie above the diagonal set of  $\mathbb{R}^2$. By default, the diagonal set of $\mathbb{R}^2$ is part of the persistence diagram, representing  all trivial homology generators that are born and die at every level. Each point on the diagonal has infinite multiplicity.  The axes of the persistence diagram are birth values on the horizontal axis and death values on the vertical axis. Again, see Figure \ref{persistence} for a schematic representation of the construction of a persistence diagram.


It is convenient to define a metric on the space of persistence
diagrams. A number of options exist. A fairly standard metric is the
$p$-Wasserstein metric. On the set of the $q$-persistence diagrams
consider the $p$-Wasserstein metric, $1\leq p\leq\infty$, defined by
\[
d_p (\mathcal{P}^1_q, \mathcal{P}^2_q) = \left(\inf_{\phi} \sum_{z \in \mathcal{P}^1_q} \| z - \phi(z) \|_{\infty}^p\right)^{1/p},
\]
where $\mathcal{P}^1_q, \mathcal{P}^2_q$ are two $q$-persistence diagrams, and  the sum is taken over all bijections $\phi : \mathcal{P}_q^1 \rightarrow \mathcal{P}^2_q$. The set of bijections, $\{ \phi:\mathcal{P}^1_q \to \mathcal{P}^2_q\}$, is nonempty owing to
the fact that each diagram includes the diagonal set, allowing one to
match off-diagonal elements in one diagram with diagonal elements in
another when their numbers differ.

The space of $q$-persistence diagrams together with the
$p$-Wasserstein metric forms a metric space, which is complete and
separable. The Wasserstein distance takes the `best' matching; that
is, it minimizes the distance, relative to the $L_p$ norm, that one
has to shift generators in $\mathcal{P}^1_q$ to match them with those
in $\mathcal{P}^2_q$. In probability theory, and in particular cases
with continuous or weighted distributions, the Wasserstein metric is
sometimes termed the `earth mover distance', which refers to the
operation of transforming one distribution into another with the
minimal change in mass. In what follows we set $p=2$ and drop the
reference to $p$. For details on the Wasserstein metric, see~\cite{Cohen-Steiner2010}.

One of the remarkable properties of persistence diagrams is their
stability, meaning that small changes in the initial point cloud data
produce persistence diagrams that are close to one another relative to
Wasserstein metric. The stability results are very general for the
`bottleneck distance', when $p=\infty$, and more restrictive for the
Wasserstein metric with $p<\infty$.  The essence of the stability
results, as shown
in~\cite{Chazal2009,Cohen-Steiner2010,EdelsbrunnerM2012}, is that the
persistence diagrams depend Lipschitz-continuously on point cloud
data.


In applications, the stability result ensures the robustness of the
data analysis performed via persistence diagrams, which makes them a
powerful alternative to statistical methods. This is particularly
useful in context of data from stochastic systems, since persistence
diagrams turn out to be quite versatile in distinguishing between
small but relevant features in a data set and noise.

\subsubsection{Detection of critical transitions via persistence diagrams} 
We now describe how to apply this method to detect critical transitions in time series. Consider a time series $(t_j,x_j)$, $j=1,\ldots, J$, with $x_j\in\mathbb{R}^d$. (In the case of a time series obtained from the model discussed in Section  \ref{section:model}, we will chose $d=1$). Assume that the  time series $(t_j,x_j)$ is obtained as a time discretization of a process $(t,x_t)$ which is Lipschitz continuous. (This is indeed the case when a time series is obtained from a Langevin equation, as in Subsection \ref{section:model}.)
To each $t_i$ we associate a   string of  $N$ consecutive data points points from the time series, with $N$ sufficiently large, which we denote
\begin{align}\label{eq:data-string}
 t_i\mapsto X_i=(x_i,x_{i+1}, \ldots, x_{i+N-1}).
\end{align}
We regard each $X_i$ as a  point cloud set in $\mathbb{R}^d$. We compute the persistence diagrams $\mathcal{P}_*(X_i)$ of $X_i$, in all dimensions, and follow the evolution  of the persistence
diagrams in time.

Diagrams corresponding to nearby times will be close to one another,
due to the Lipschitz continuity of the process underlying the time
series and to the robustness of persistence diagrams. Within this context,
persistence diagrams that are near to each other in time, but
relatively far from one another in the Wasserstein metric, indicate a
sudden change in the time series. Therefore, we propose the following
empirical criterion for detection of critical transitions in slow-fast
systems:
\begin{itemize}
  \renewcommand{\labelitemi}{$\vcenter{\hbox{\tiny$\bullet$}}$}
  \item Persistence diagrams undergo significant changes, measured
    using the Wasserstein metric, prior to a critical transition.
\end{itemize}

This criterion follows from the following heuristic argument.
If the noise level
in the Langevin equation is small, then far from a critical transition
the time series follows closely, with high probability, a trajectory
of the slow subsystem. A point cloud associated to a data string
displays significant topological features similar to those of the slow
manifold, plus less significant topological features due to noise.
In addition, the corresponding persistence diagrams at nearby times are close
to one another relative to the Wasserstein metric.  

When the system undergoes a critical transition, the time series
ceases to follow the slow manifold, as the dynamics enters a transient
regime. The topological features associated to the slow manifold are
destroyed, and new topological features appear in the point cloud
structure. Furthermore, if the system moves to a different stable
regime after a finite time, the point cloud will reflect the
topological features associated to that regime. Critically, for a
point cloud data {\em near} a critical transition the corresponding
persistence diagrams shift away from those diagrams corresponding to
data far from the critical transition.  Consequently, successive
distances between diagrams in this region exhibit a large jump prior
to a critical transition.


\section{Results}\label{sec:results}
\begin{figure}\centering
\includegraphics[width=0.7\textwidth, clip, keepaspectratio]{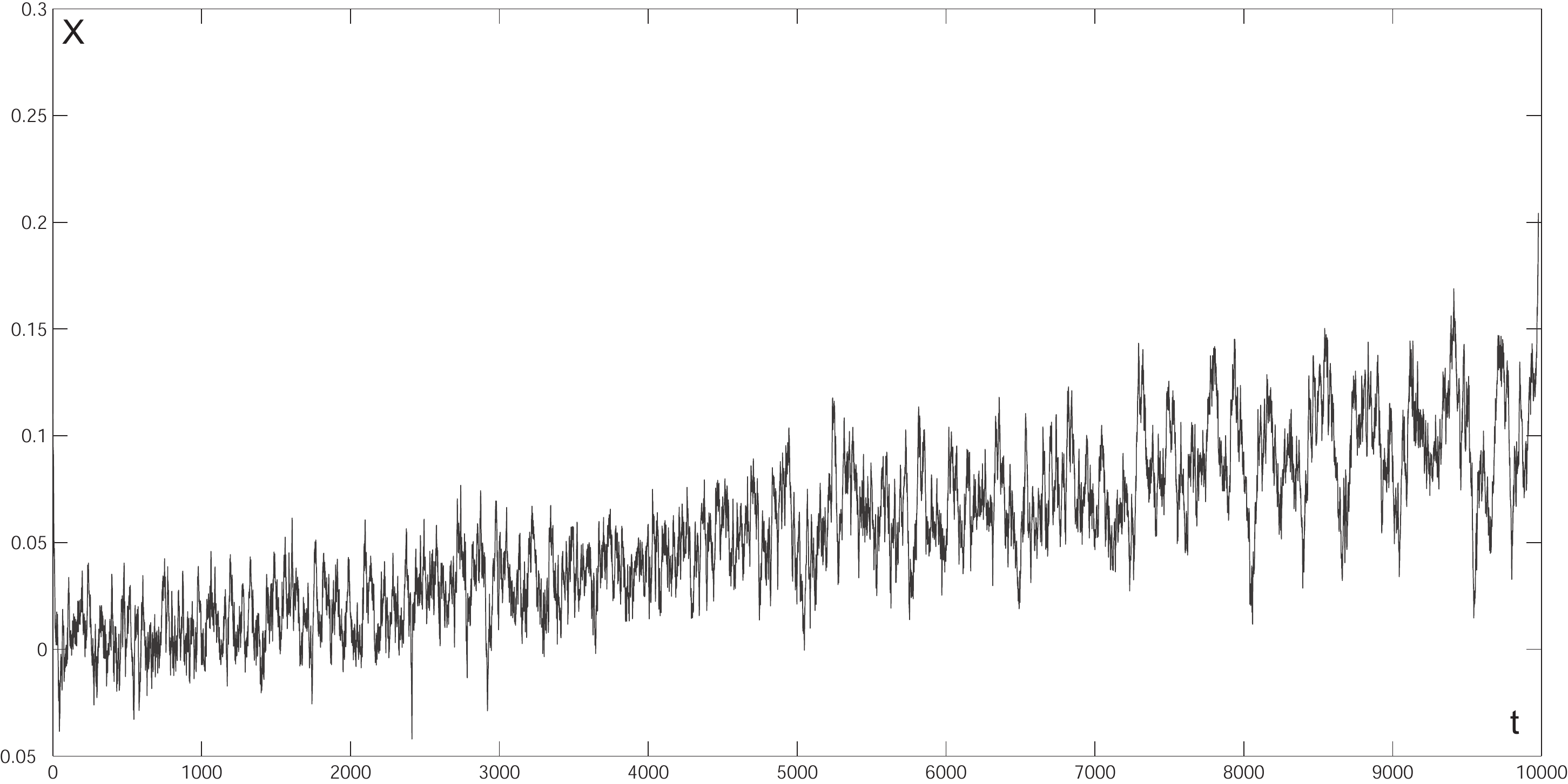}
\caption[]{$R$-time series for the gene regulatory network model from Section \ref{section:model} cut off before the critical transition.}
    \label{cutoff}
\end{figure}

We numerically solve the SDE defined in \eqref{eqn:Xr_stoch} -- \eqref{eqn:Sr_stoch} using the Euler-Maruyama procedure, with stepsize $0.01$ and noise level $\sigma_1=\sigma_2=0.02$. We fix the rate of change of the parameter $S$ to be $\eps=10^{-4}$.
As output, we choose the time series given by the $R$-component; a particular realization of this time series is shown in the righthand panel of Fig. \ref{substratedepletion}. Since the solution values are dense in time, we subsample the time series by taking every $10$-th data point.
The time series follows a slowly varying attractive equilibrium point, until it reaches a critical transition, at which point it enters an oscillatory mode. To test for early signs of the critical transition, we  truncate the time series before it enters the oscillatory regime. This truncated region is shown in Fig.~\ref{cutoff}.

\subsection{DFA analysis}\label{subsection:DFAanalysis}
\begin{figure}\centering
\includegraphics[width=1.0\textwidth, clip, keepaspectratio]{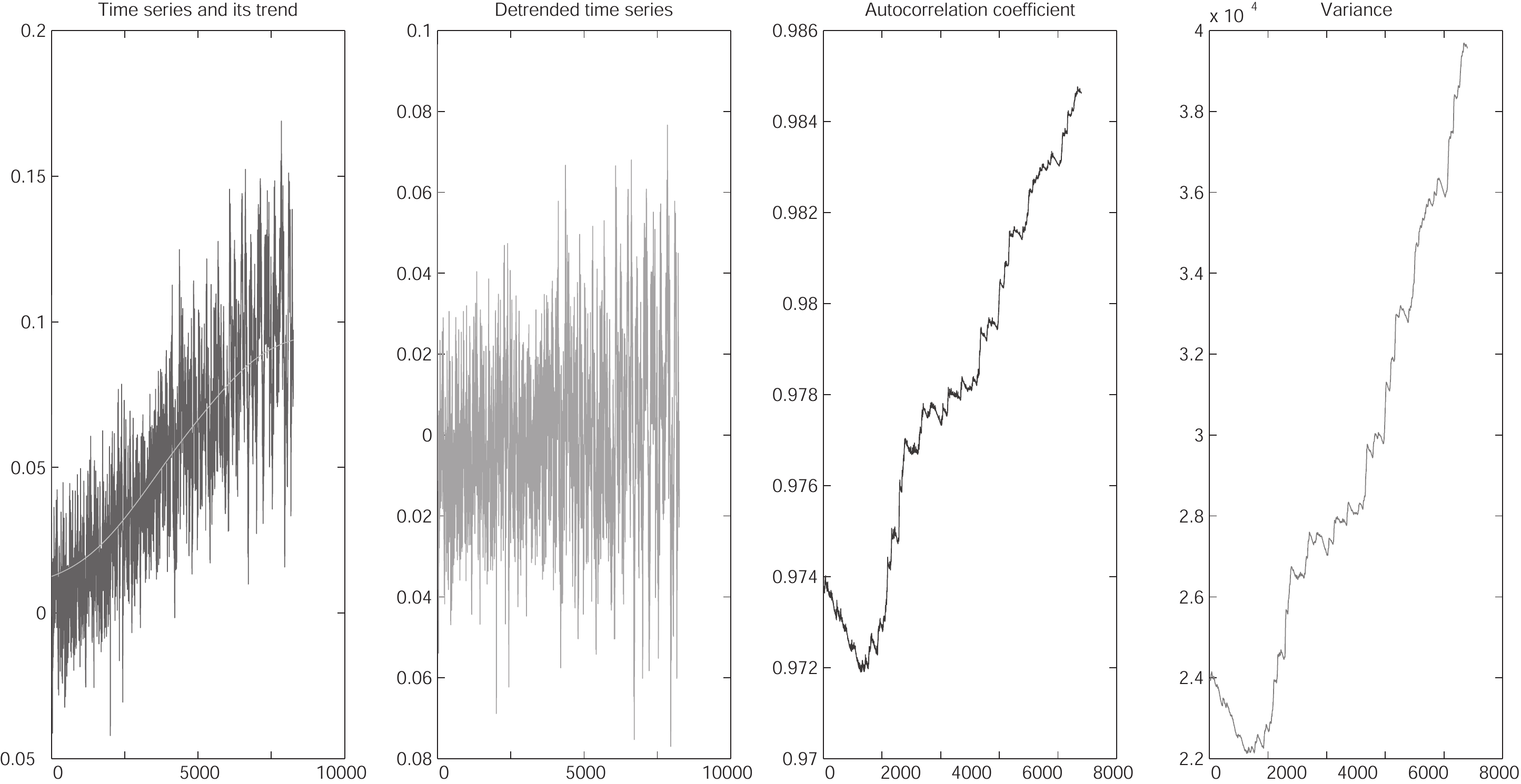}
\\
\includegraphics[width=1.0\textwidth, clip, keepaspectratio]{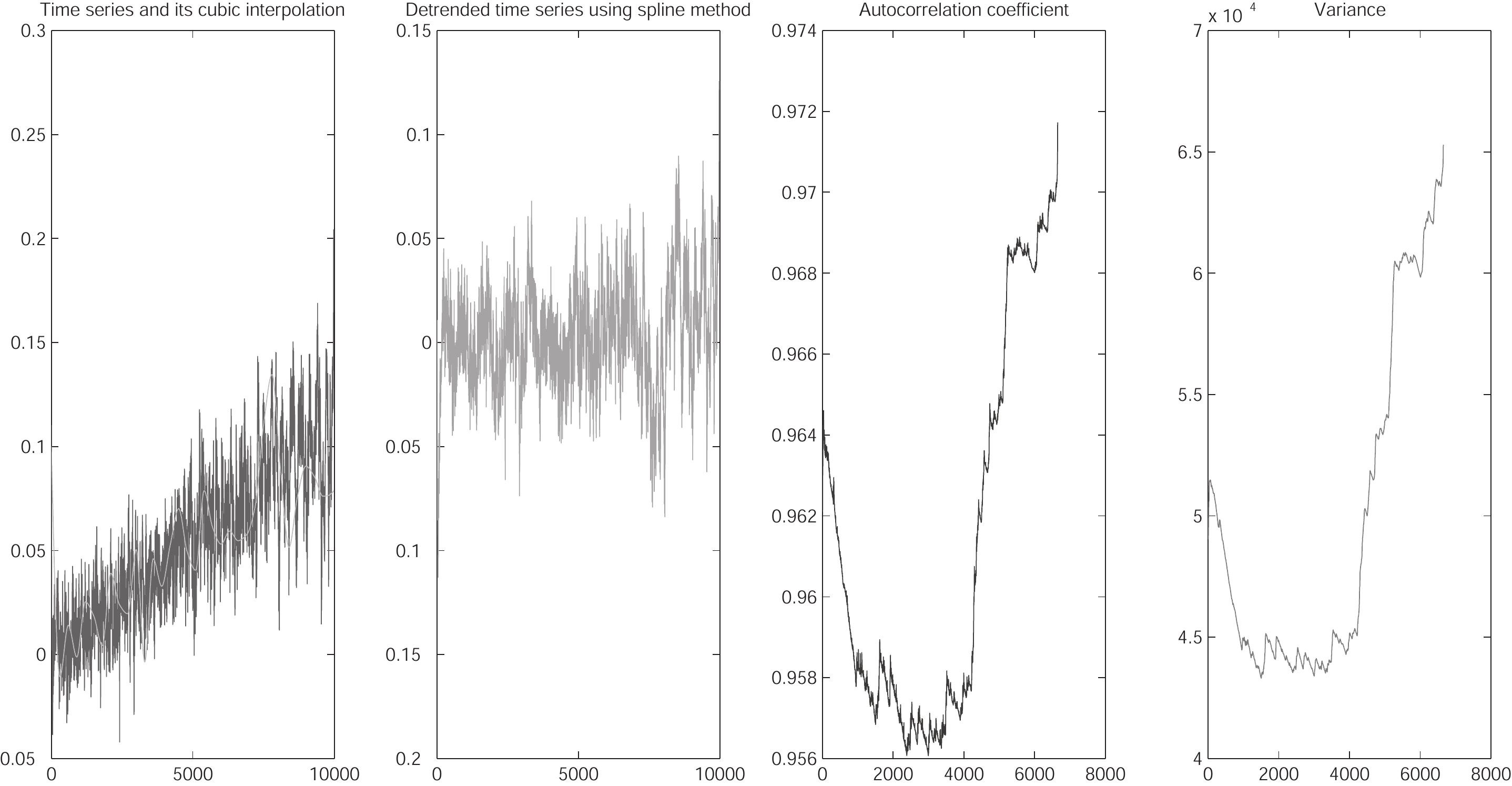}\
\includegraphics[width=1.0\textwidth, clip, keepaspectratio]{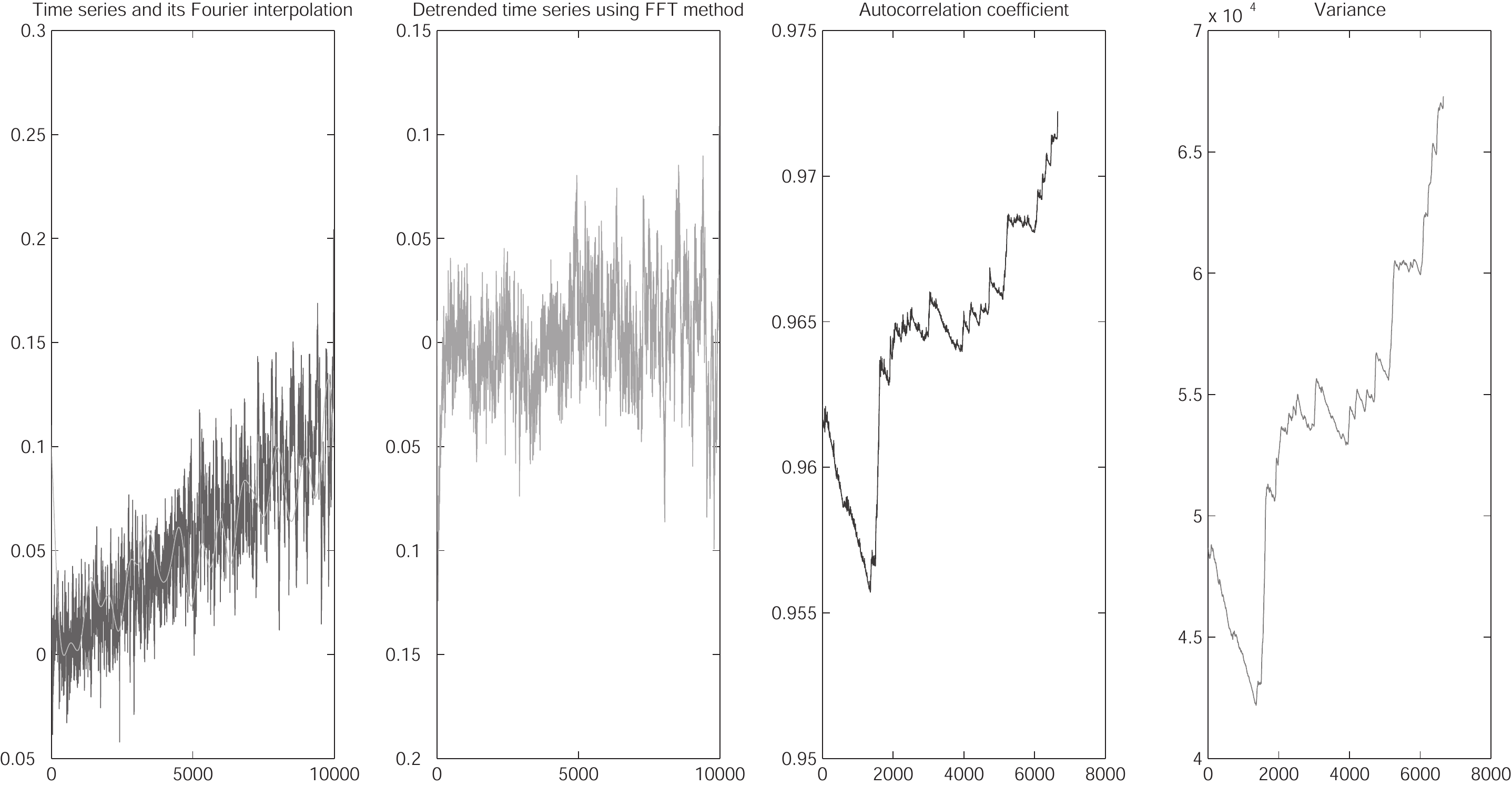}
\caption[]{DFA analysis of the time series using (top to bottom) Gaussian kernel, cubic spline, and Fourier interpolation detrending.}
    \label{substratedepletionDFA}
\end{figure}

We conduct three experiments  to detect critical transitions using the DFA\footnote{The DFA procedure that we use here was implemented in Matlab  by Rebecca M. Jones \cite{Jones}}  methodology on the time series generated by our model.
The first experiment uses a Gaussian kernel to detrend the time
series; in the second experiment we use a cubic spline interpolation;
and in the third experiment, we use Fourier interpolation. For each of
the detrended time series we compute $AR(1)$ and the variance for a windowed time series. The
results are summarized in Fig. \ref{substratedepletionDFA}. All three
experiments show $AR(1)$ increases to $1$ as the system approaches the
transition, while the variance also grows steadily, both behaviors
being consistent with a critical transition.

\subsection{Persistence diagram analysis}\label{subsection:persistenceanalysis}
We compute the $0$-dimensional persistence diagrams for strings of $N$
data points $X_i$ (see Eq.~\eqref{eq:data-string}) as the $t_i$ approaches
a critical transition, as for the example time series in Fig.~\ref{cutoff}. We construct Rips complexes,
with an initial radius of $\delta=10^{-4}$ around each data point, and
then grow the radii of the balls $\delta,2\delta,\ldots,
n\delta$, where $n$ is chosen large enough so that the final complex
in the resulting filtration has a single connected component. From
this filtration we compute the $0$-persistence diagrams. We choose the
size of the data sets $N=300$; larger sizes make little difference in
the qualitative behavior of the diagrams.

The $0$-dimensional persistence diagrams are easy to interpret: they
track of the births and deaths of connected components in the Rips
complex, as the radii of the balls increase.  Note that all births
occur at the same time, when the radius of balls is zero and we have
$N$ disjoint points. After this initial stage, a large
number of connected components die as the radii of the balls
increase, as connected components merge with one another.

\begin{figure}\centering$\begin{array}{cc}
\includegraphics[width=0.45\textwidth, clip, keepaspectratio]{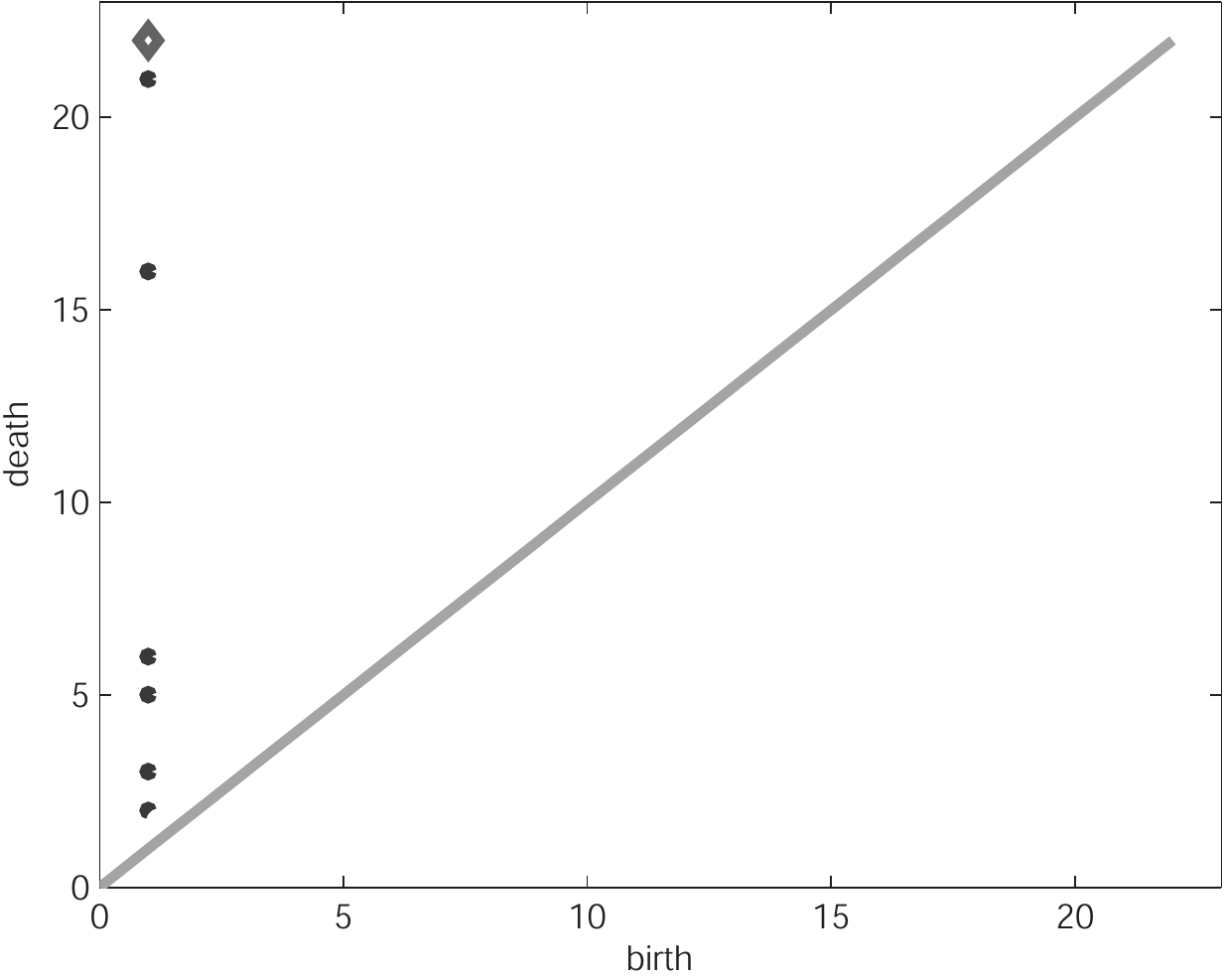} &  \includegraphics[width=0.45\textwidth, clip, keepaspectratio]{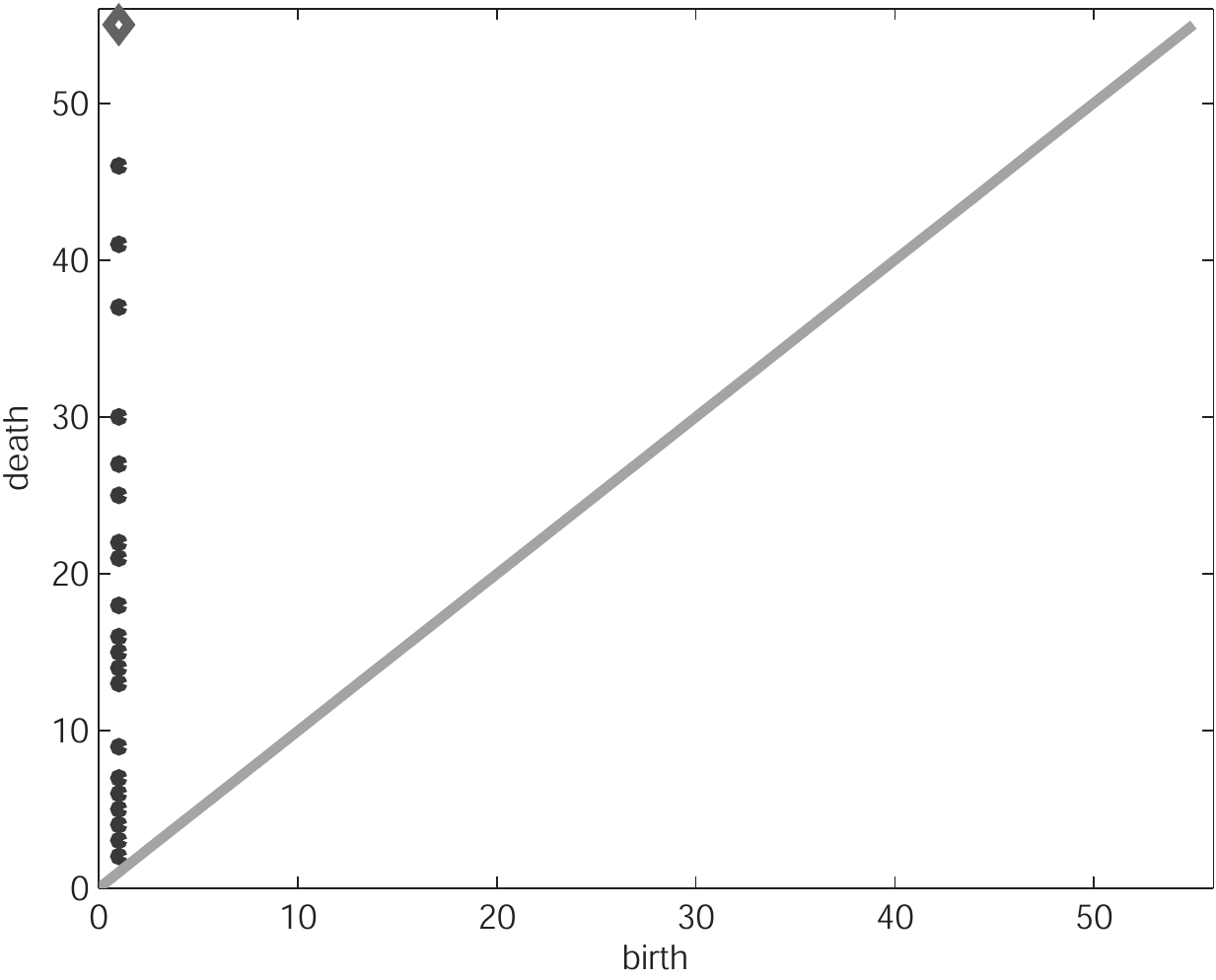}
\end{array}$
\caption[]{The 0-persistence diagram on the left corresponds to a
  string of data sampled far from the critical transition; the diagram on
  the right corresponds to a string of data sampled near the
  critical transition. In each persistence diagram, the dots mark
  finite deaths times, and the diamond indicates infinite death
  times.}\label{fig:persistence}
\end{figure}

Consider Fig.~\ref{fig:persistence}, where the persistence diagrams for
a data string far from the critical transition (left panel) and a data
string close to the critical transition (right panel) are shown.  For
data far from the critical transition, points cluster near the
attractive equilibrium. Thus, when balls are constructed in the Rips
filtration around these points, they will quickly yield a robust
connected component around the attractive equilibrium , plus a
small number of scattered connected components corresponding to points
that escape for brief periods time from the  equilibrium
point due to stochastic effects.
The implication is that, in the
corresponding persistence diagram, the vertical spread of the death
times is relatively small, and consists of a small numbers of points
away from the diagonal (accounting for the robust connected component
and a few outliers), plus many short-lived points close to the diagonal
(accounting for noise).  

Conversely, when a data string originates from close to a critical
transition, the points tend to spread further away from the attractive
equilibrium point, due to changes in the potential
field. Heuristically, the equilibrium loses its
attractiveness. This causes 
 the distribution of the
data points from the time series to grow. The implication is
that, in the corresponding persistence diagram, the vertical spread of
the death times is much larger, with a tendency to form multiple small
clusters.

\begin{figure}
\centering
\includegraphics[width=1.05\textwidth]{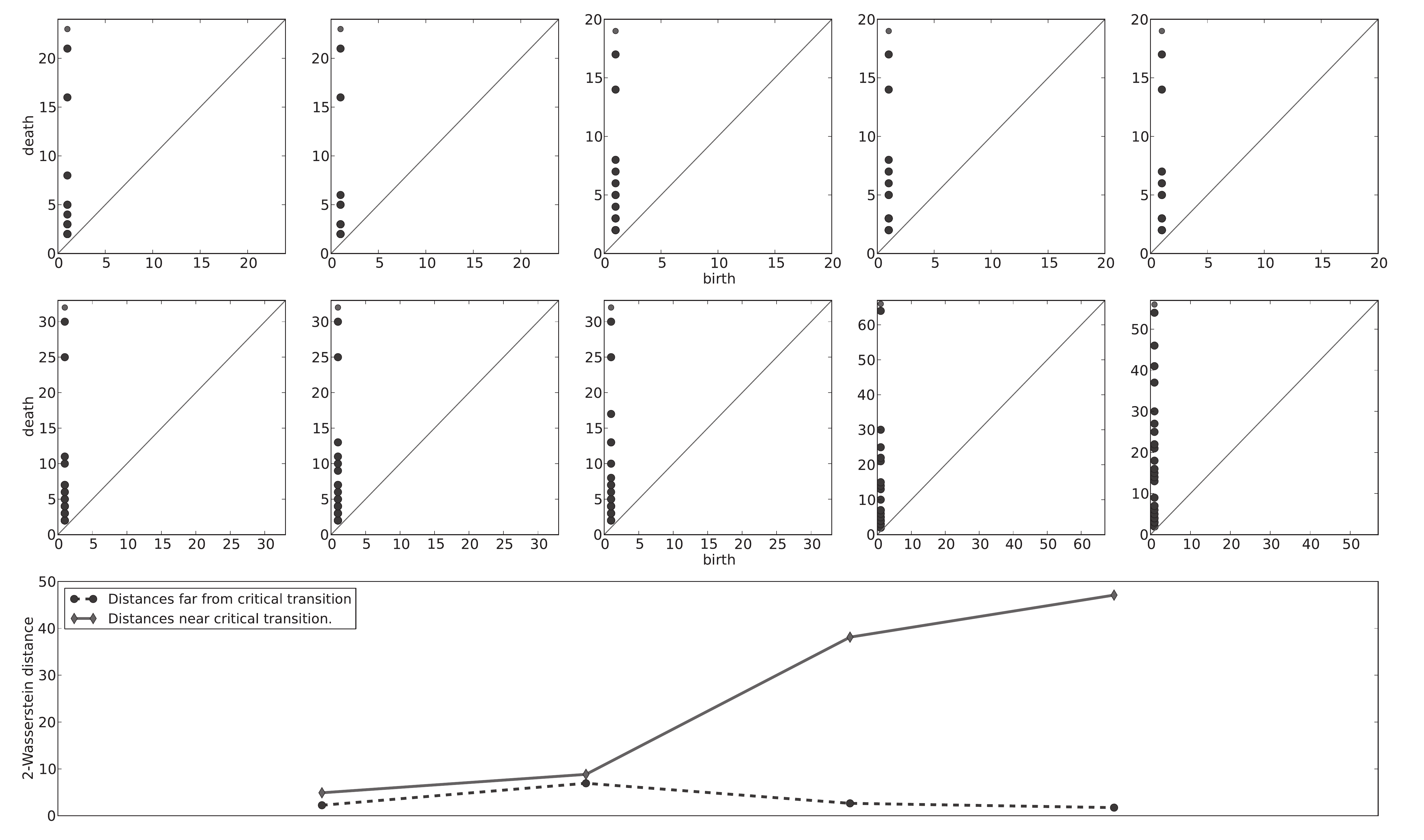}
\caption[]{The $0$-persistence diagrams on the top row correspond to
  consecutive strings of data sampled far from the critical
  transition; the diagrams in the middle row correspond to consecutive
  strings of data sampled near the critical transition. The bottom
  curves describe the distances between consecutive diagrams. The plot
  labels are positioned between two successive diagrams with values on
  the $y$-axis indicating the 2-Wasserstein distances between
  them.}\label{fig:sd}
\end{figure}

The visual inspection of persistence diagrams provides intuition, but
is not a precise way to indicate the approach to a critical
transition. To quantify the above assessment, we study the behavior of
the Wasserstein distances between consecutive diagrams which are
summarized in Fig.~\ref{fig:sd}. In the figure, time increases from
left to right. The figures in the top row represent persistence
diagrams for data sampled far from the critical transition, and those
in the middle represent persistence diagrams for data sampled close to
the critical transition. The five time frames captured in each column
spread over a time interval of size $\Delta t=0.5$.  We then compute
the Wasserstein distances between consecutive persistence
diagrams. 
These changes in the persistence diagrams are quantified in the bottom
row of Fig.~\ref{fig:sd}.  The solid curve, corresponding to data
near the critical transition, shows a significant increase in the
distances between consecutive diagrams as the point cloud anayzed near
the critical transition.  The dotted curve, corresponding to distances
between diagrams far from the critical transition, shows only small
variations in the consecutive distances. The computed distance are
indicated by the symbols on each curve and are placed between the
diagrams from which they were computed.\footnote{For the computation
  of persistence diagrams we use the Perseus software developed by
  Vidit Nanda~\cite{Nanda}. The Wasserstein distances were computed
  using software written by Miroslav Kramar~\cite{Miro2013}.}

Note that the variance of the lifespans is related to the variance in
the time series. An increase in the vertical spread while approaching
a critical transition is consistent with the findings by the DFA
method in Subsection~\ref{subsection:DFAanalysis}. Also, the change observed in
the clustering of the diagram coordinates is related to asymmetric or
multimodal properties of the data, as mentioned in
Subsection~\ref{subsection:persistence}.


\section{Conclusions}

The substrate-depletion oscillator that we analyze in this paper is a
realistic model for certain types of molecular regulation circuits
studied experimentally. The methods for detecting critical transitions
that we propose are suitable for the analysis of real data as
well. Indeed, most of the experimental data obtained about gene
regulatory networks (e.g., data obtained from microarrays, or reverse
transcriptase polymerase chain reaction) is limited by background
noise, and both the DFA and persistence diagram methods are robust to
noise in data, as long as the noise does not overwhelm the
signal. Also, in comparison with the DFA method, which necessitates a
number of `ad-hoc' choices of statistical parameters and procedures,
the persistence diagram method appears more robust and
objective.

In current and future work, we are developing a theoretical framework for the
empirical criterion proposed in this paper.  Namely, we plan to
establish rigorously that bifurcation-induced critical transitions
determine large changes in the persistence diagrams, and, conversely,
large changes in the persistence diagrams imply the existence of
bifurcations.

\section*{Acknowledgement}
A portion of this work has been done while  M.G. was a member of the IAS, to which he is  very grateful. Also, we thank Konstantin Mischaikow, Miroslav Kramar, Vidit Nanda, and Rebecca M. Jones  for many useful discussions on this subject.



\begin{thebibliography}{XXXXX}

\bibitem{Bulter2004} T. Bulter,  S.-G. Lee,   W.-W. Wong,  E. Fung,   M.R. Connor,   and J.C. Liao,  Design of artificial cell-cell communication using gene and metabolic networks, Proc. Natl. Acad. Sci. USA, 101 (2004), 2299--2304.

\bibitem{Bryce2012} R.M. Bryce and K.B.  Sprague,  Revisiting detrended fluctuation analysis, Scientific Reports 2, (2012),  doi:10.1038/srep00315

\bibitem{Chazal2009} F. Chazal, D. Cohen-Steiner, L. J. Guibas, F. M\'{e}moli, S. Oudot, Gromov-Hausdorff Stable Signatures for Shapes using Persistence, Computer Graphics Forum (proc. SGP 2009)    (2009), 1393--1403.

\bibitem{Chazal2011} F. Chazal, L.Guibas, S. Oudot, and  P. Skraba, Persistence-Based Clustering in Riemannian Manifolds, Proc. 27th Annual ACM Symposium on Computational Geometry,   (2011), 97--106.
    
\bibitem{Chazal2012} F. Chazal, V. de Silva, S. Oudot,  Persistence stability for geometric complexes, Geometriae Dedicata, (2013), doi: 10.1007/s10711-013-9937-z

\bibitem{Cohen-Steiner2010} D. Cohen-Steiner, H. Edelsbrunner, J. Harer, Y. Mileyko.
Lipschitz Functions Have $L_p$-Stable Persistence. Foundations of Computational Mathematics. Vol. 10 (2010), doi: 10.1007/s10208-010-9060-6

 

\bibitem{Ditlevsen2010} P.D. Ditlevsen and S.J. Johnsen, Tipping points: Early warning and wishful thinking,
Geophysical Research Letters, Vol. 37, L19703  (2010),  doi:10.1029/2010GL044486

\bibitem{Edelsbrunner2008} H. Edelsbrunner  and  J. Harer, Persistent homology --- a survey, in ``Surveys on Discrete and Computational Geometry. Twenty Years Later", 257-282, eds. J. E. Goodman, J. Pach and R. Pollack, Contemporary Mathematics 453, Amer. Math. Soc., Providence, Rhode Island, (2008).

\bibitem{EdelsbrunnerM2012} H. Edelsbrunner and M.   Morozov, Persistent Homology: Theory and Applications,
Proceedings of the European Congress of Mathematics, (2012).




\bibitem{Elowitz}    M.  Elowitz and S.Leibler,  A Synthetic Oscillatory Network of Transcriptional Regulators,  Nature,  Vol. 403,  (2000), 335--338.

\bibitem{Fenichel79} N.  Fenichel, Geometric singular perturbation theory for ordinary differential equations, Journal
of Differential Equations, 31  (1979), 53--98.

\bibitem{Gameiro2012} M. Gameiro, Y. Hiraoka,  S. Izumi, M. Kramar, K. Mischaikow,  and V. Nanda,   A Topological Measurement of Protein Compressibility, preprint, (2013).

\bibitem{Gardner}  T.S Gardner,  C.R.  Cantor,  and J.J. Collins,  Construction of a genetic toggle
switch in Escherichia coli,  Nature, Vol. 403 (2000), 339--342.


\bibitem{Hasty2000}  J. Hasty, J.  Pradines, M. Dolnik,
 and J.J. Collins,   Noise-based switches and amplifiers for
gene expression,  PNAS 97 (2000), 2075--2080.


\bibitem{Hatcher2002}  A. Hatcher,{\em Algebraic Topology}. Cambridge University Press, (2002).

\bibitem{Jones} R.M. Jones, Matlab code for the DFA procedure,
http://criticaltransitions.wikispot.org/ (2013).

\bibitem{Kuehn2011} C. Kuehn, A mathematical framework for critical transitions: Bifurcations, fast-slow systems and stochastic dynamics, Physica D: Nonlinear Phenomena, 240(12) (2011), 1020–-1035.

\bibitem{Kuehn2012} C. Kuehn, A Mathematical Framework for Critical Transitions: Normal Forms, Variance and Applications, Journal of Nonlinear Science, DOI 10.1007/s00332-012-9158-x

\bibitem{KrupaS2001}  M. Krupa, and  P. Szmolyan, Extending geometric singular perturbation theory to nonhyperbolic points -fold and canard points in two dimensions, SIAM J. of Math. Anal. 33 (2001), 286-- 314.

\bibitem{Kondic2012}  L. Kondic,  A. Goullet, C.S. O'Hern,   M. Kramar,   K.Mischaikow,   R.P. Behringer,
Topology of force networks in compressed granular media,
Europhys. Lett. 97 (2012), 54001.

\bibitem{Miro2013} M. Kramar, C++ code to compute the Wasserstein metric, Personal communication, http://www.math.rutgers.edu/~miroslav (2013).

\bibitem{Leier2006}  A. Leier, P.D. Kuo,  W. Banzhaf, and  K. Burrage,  Evolving noisy oscillatory
dynamics in genetic regulatory networks, In  EuroGP'06 Proceedings of the 9th European conference on Genetic Programming, LNCS 3905 (2006), 290--299.

\bibitem{Livnia2007} V.N. Livina  and T. M. Lenton,  A modified method for detecting incipient bifurcations in a dynamical system, Geophys. Res. Lett., 34,  (2007), L03712.



\bibitem{Nanda}  V. Nanda, The Perseus Software Project for Rapid Computation of Persistent Homology, http://www.math.rutgers.edu/\~{}vidit/perseus.html

\bibitem{Nicolau2011} M. Nicolau, A.J. Levine,  and G. Carlsson, Topology based data analysis identifies a subgroup of breast cancers with a unique mutational profile and excellent survival, PNAS,17 (2011)

\bibitem{Novak1993} B. Novak, and J.J. Tyson, Numerical analysis of a comprehensive
model of M-phase control in Xenopus oocyte extracts and
intact embryos, J. Cell. Sci., 106  (1993), 1153--1168.

\bibitem{Ptitsyn2007} A.A. Ptitsyn, S. Zvonic, and J.M. Gimble,  Digital Signal Processing Reveals
Circadian Baseline Oscillation in Majority of Mammalian Genes, PLoS Comput
Biol 3  (2007), e120.




\bibitem{Scheffer2009} M. Scheffer et al., Early-warning signals for critical transitions,
Nature, 461 (2009), 53--59.

\bibitem{Scheffer2012} M. Scheffer et al, Anticipating Critical Transitions, Science 338, (2012) 344, 
DOI: 10.1126/science.1225244




\bibitem{Thompson2010} J.M.T. Thompson, and J.  Sieber, Climate tipping as a noisy bifurcation: a predictive technique, IMA Journal of Applied Mathematics (2010), 1--20.

\bibitem{Tyson2003} J.J. Tyson,  K.C. Chen, and  B.  Novak,  Sniffers, buzzers, toggles and blinkers: dynamics of regulatory and signaling pathways in the cell,  Curr. Opin. Cell. Biol. 15(2),  (2003), 221--231.

\bibitem{Zomorodian2005} A.J. Zomorodian, Topology for Computing, Cambridge University, (2005)


\end{thebibliography}
\end{document}